# Some Fundamental Properties of the Integer-States in Open-System, Ensemble Energy-Density Functional Theories

Steven M. Valone
Materials Science and Technology Division, Los Alamos
National Laboratory, Los Alamos, NM 87545

*Ensemble averages are an approximation technique for connecting macroscopic and microscopic properties of a system. For systems open with respect to exchange of particles with a bath, the microscopic states are those with integer numbers of particles. When a property of the open system is represented as an ensemble average over these microscopic states, self-consistency dictates several implications for the properties of both for open-system energy density functionals and the integer-state functionals describing the microscopic states. The first is that each integer-state energy density functional is a functional of the original type discovered by Levy. Another is that the dependence of the open-system functional on the ensemble density is linear whereas the dependence of the integer-state functional is decidedly nonlinear. Finally, the derivative discontinuity behavior with respect to particle number of some open-system density functionals appears to be connected to the long-range behavior of the effective external potentials of the integer-state functionals governing the interactions among subsystems in the ensemble.*

# I. Introduction

The behavior of complex materials, the foundations for modern technology, presents huge challenges for the modeling and theory communities. One strategy for addressing complexity in materials behavior is through the implementation of open-system ensemble formalisms. In this context, by an open system, we mean one whose particle number varies with time through exchange of particles with a bath. In spite of the long interest in this approach, questions still abound over the behavior of open ensembles especially when applied in an energy density-functional context.[1,2] The best known example of such an application is the work of Perdew et al.,[3] hereafter referred to as PPLB. There the behavior of the derivative the energy with respect to the average number of particles is examined in detail. The implications of the discontinuous behavior of this quantity, first discussed in detail by Gyftopoulos and Hatsopoulos[4], are followed to their logical conclusion for the functional derivative of the energy with respect to the average density.[3] In this work, we will focus on some properties of the individual states composing the ensemble, rather than the focusing on the average values. For this purpose, imagining the possible states of an atom embedded in a molecule or crystal captures that focus.

By design open ensemble formalisms assume very little knowledge about a physical system. Even the number of particles in the microscopic states composing the ensemble, much less the wavefunction for the total system, are not completely specified. For this reason open ensemble approaches are very useful at the same time that they suffer from ambiguity. In one formulation by Reif[5] the ambiguity can be viewed as stemming for ones choice about which properties of the ensemble are known. One may assume knowledge of bath characteristics pertaining to the macroscopic state or one may assume knowledge of microscopic states and their time averages. (Time and statistical averages are assumed to be identical here.) It is only when the ensemble is a thermodynamic-equilibrium one,[5] that one can assume that knowledge of macroscopic and microscopic quantities are interchangeable.

In general, macroscopic or microscopic properties can be assumed to be either known or unknown. Given known properties at either level, one deduces unknown ones from them but with fluctuations occurring about the average values. Under special circumstances, the fluctuations are small enough to ignore. However, the "smallness" depends on the magnitude of the interactions between the two subsystems. Obviously then, the "smallness" determination must be made on a case-by-case basis. The affect of the constraints imposed by known properties on the unknown ones is also an integral part of this determination. In the formal development that follows, issues of "smallness" are discussed as little as possible. The reader should keep in mind, though, that they are still present.

As an example, a typical situation consists of an open system in contact with a bath at temperature T. Following Gyftopoulos and Hatsopoulos[4], the average energy, *E*, and average number of particles, *N*, given by the summations

$$E = \sum_q W_q \mathbf{E}_q \,, \tag{1}$$

and

$$N = \Sigma_q \, W_q \, \mathbf{N_q} \,, \tag{2}$$

over the states $0 \leq q \leq S$, subject to the constraints that each occupation number $W_q \geq 0$ and

$$1 = \Sigma_q \, W_q \,. \tag{3}$$

Bold-faced quantities $\mathbf{E_s}$ and $\mathbf{N_s}$ are the energies and numbers of particles in the microscopic states, "s".  These microscopic states will be referred to as integer states.  (Here and below italicized variables will always represent time-averaged quantities and bold-faced variables will always represent integer-state properties.)

In turn, since we are assuming that the system is a thermodynamic one, the occupation numbers are related to the temperature, T, and the chemical potential, $\mu$, through the ratio

$$W_q = w_q / \Sigma_q \, w_q \,, \tag{4}$$

where

$$w_q = \exp((\mu \, \mathbf{N_q} - \mathbf{E_q})/k_B T) \,, \tag{5}$$

with $k_B$ being the Boltzmann constant.  Eq. (5) ties the time-averaged variables to the stationary variables.  Expressed this way normalization, Eq. (3), is automatically satisfied.

Typically, both T and $\mu$ are viewed as a properties of the bath.  However, as pointed out by Reif,[5] $\mu$ can be viewed as a property of either the bath or as a chemical potential of a microscopic system.  If, as assumed in Gyftopoulos and Hatsopoulos[4] and PPLB[3], $N$ is known for a microscopic system of interest, μ is a chemical potential of that system as determined by Eq. (2), but not necessarily of the bath. All of the conclusions of Gyftopoulos and Hatsopoulos[4] on the behavior of open-system ensemble averages and of PPLB[3] for the energy density functionals derived therefrom hold. If one assumes that the chemical potential of the bath is known, then $N$ is determined as a function of μ by Eq. (2). In this case, it is reasonable to ask what the behavior of $\partial E/\partial N$ is when μ of the bath is known.

The goals of the present investigate are: (1) To construct an open-system, ensemble average of energy density-functionals for atomic and molecular systems governed by the Hohenberg-Kohn theorem[6] (DFT) and constrained-search density-functional theory[7,8] (CS-DFT); (2) To understand the properties of the integer-states of the ensemble, given properties of the averages and/or the ensemble; (3) To relate the derivative discontinuity of the time-averaged exchange-correlation energy density-functional with respect to the time-averaged particle number to the analogous integer-state functional derivatives; and (4) Describe an iterative procedure for achieving consistency in the properties between or among the subsystems of the closed, isolated system in a manner suggested by Rychlewski and Parr[9]. The seminal work of PPLB,[3] shows that both formalisms apply to systems open to particle exchange. Constructing the ensemble averages will, following Reif,[5] require reversing the normal order in which assumptions are adopted. Because of the ambiguous nature of open-system ensembles, there are several points at which somewhat arbitrary choices are required. In this sense, the results here are not completely general. They are, however, more general than those present in standard texts on thermodynamics.  In the next section, at least a subset of the possible ensemble averages of energy density-functionals will be constructed through a well-defined set of assumptions and approximations.  In the subsequent section, some properties of the integer-states are explored.  In particular, it will be found that, for this subset of ensemble averages, (1) Each integer-state contributing to the ensemble average is an integer-state energy density-functional in the sense of Levy[7] and Valone[8];  (2) Chemical potential equalization between a reservoir and an open system is passed on to each integer-state in the ensemble average;  (3) The coefficients for averages of the chemical potentials are determined by a time-superposition principle, when the time-averaged number of particles is specified, rather than the chemical potential being specified;  and (4) The case where the ensemble is a thermodynamic one are consistent with the general case.  The implications for these results have a bearing on how energy-density functional calculations are performed, the construction of empirical potential energy surfaces, and the construction of new energy-density functionals.

## II. Construction of an ensemble average

Our goal in this section is to derive an ensemble average starting, as does Reif[5], from a closed system in its quantum mechanical ground state that will encompass both the bath and the microscopic system of interest.  However, the appropriate pathway for our derivation adopts assumptions only as needed in order to derive an ensemble average for which CS-DFT is applicable.  So, in the beginning, no reference will be made to baths and microscopic subsystems.

For our purposes, the closed, isolated system is a fixed configuration of atoms in a molecule, liquid or solid.  The hamiltonian H for the system has N electrons, ground state $\Psi$, energy E, and subsystems $H_r$ and $H_s$, such that $H_r + H_s + V_{rs} = H$.  The partitioning corresponds to assignment of atomic centers to one subsystem or another, but precise details of the subdivision are left unspecified here, as is the interaction potential, $V_{rs}$.  The details depend on which properties are most desired.  In particular, at this point, no specification is made as to the relative sizes or stabilities of the two subsystems.

Now consider the combinations of states which the two subsystems can be in at any instant of time, including differing numbers of electrons but at constant total number of electrons.  We add the index q to indicate the partitioning of electrons between the two subsystems.  Let $H_{rq}$ have $N_{rq}$ electrons which requires that $H_{sq}$ have $N_{sq} = N - N_{rq}$ electrons.  In addition, let the subsystem r wavefunctions in quantum mechanical state q be $\psi_q$ with energy $E_{rq}$ such that

$$H_{rq} \psi_q = E_{rq} \psi_q . \tag{6}$$

To this point, the interaction Hamiltonian, $V_{rs}$, has not been included to any degree, and the states of subsystem s, $\phi_q$, are unspecified.  Our intent is to construct an ensemble average governing these wavefunctions and to define a density function from this average.

Next consider the usual expression for the expectation value for the energy of the total system:

$$E = \langle\Psi|H|\Psi\rangle/\langle\Psi|\Psi\rangle,  \quad (7)$$

where the angular brackets $\langle\ldots\rangle$ indicate integration over all N spin-space coordinates, $(\mathbf{r},\sigma)$. For this point on, the states of $H_r$ can be used to define the states of $H_s + V_{rs}$ and vice versa. To achieve this, perform an average over the subsystem $H_{rq}$ when it has $N_{rq}$ electrons and sum over all such possibilities. The expansion is performed over both $\Psi$ and $\Psi^*$, its complex conjugate. The resulting approximation is:

$$E \approx \left(\Sigma_{q,q'} \langle\langle\Psi|\psi_{q'}\rangle_{rq'} \langle\psi_{q'}|H|\psi_q\rangle_{rq} \langle\psi_q|\Psi\rangle_{rq}\rangle_{sq}\right)$$

$$/\left(\Sigma_q \langle\langle\Psi|\psi_q\rangle_{rq} \langle\psi_q|\Psi\rangle_{rq}\rangle_{sq}\right), \quad (8)$$

where the angular brackets $\langle\ldots\rangle_a$ indicate integration over all $N_a$ spin-space coordinates for "a" equal to either "rq" or "sq" coordinates. Notice that the integrations only make sense if $N_{sq}$ and $N_{sq'}$ are equal. No claims of completeness of the subsystem states $\psi_q$ is made at this point. Hence we indicate the relationship between E and the summation as an approximate one. Also spin exchange between subsystems has been ignored, although each subsystem state and the total system wavefunction are assumed to be fully antisymmetrized. Finally, note that $\psi_q$ need not be a ground-state wavefunction.

It is now possible to define another set of wavefunctions for the subsystem s in state q, $\phi_q$, based on the subaverages of $\Psi$ over the subsystem $\psi_q$ and its coordinates:

$$\phi_q = <\psi_q|\Psi>_{rq} . \tag{9}$$

Definitions for occupation numbers $w_q$ follow natural as:

$$w_q = < <\Psi|\psi_q>_{rq}<\psi_q|\Psi>_{rq} >_{sq} = <\phi_q|\phi_q>_{sq} . \tag{10}$$

The wavefunctions $\phi_q$ are not necessary normalized but do inherit antisymmetry from $\Psi$. Next we take advantage of the fact that $\psi_q$ is an eigenstate and only involves $N_{rq}$ electrons, say electrons $1,\ldots,N_{rq}$. Further, the $\psi_q$ are taken to be orthonormal. Also, if $N_{rq}$ and $N_{rq'}$ are different, then $\psi_q$ and $\psi_{q'}$ are assumed to be orthgonal. The approximate expression for E becomes

$$E \approx \left(\Sigma_q w_q (E_{rq} + <\phi_q|H_s|\phi_q>_{sq}/<\phi_q|\phi_q>_{sq}) + \right.$$

$$\Sigma_{q,q'} (w_q w_{q'})^{1/2} <<\phi_{q'} \psi_{q'}|V_{rsq}|\psi_q \phi_q>>)/(<\phi_{q'}|\phi_{q'}>_{sq'} <\phi_q|\phi_q>_{sq})^{1/2})$$

$$/\left(\Sigma_q w_q\right) . \tag{11}$$

We have reached a pivotal point in the construction. A separate assumption is necessary to achieve the desired ensemble average. First one can simply make the formal definition of an effective interaction-potential contribution to the external potential, as

$$<\phi_q|V_{rsq}^{ext}|\phi_q>_{sq}/<\phi_q|\phi_q>_{sq} \equiv$$

$$\Sigma_{q'} (w_q w_{q'})^{1/2} <<\phi_{q'} \psi_{q'}|V_{rsq}|\psi_q \phi_q>>)/(<\phi_{q'}|\phi_{q'}>_{sq'} <\phi_q|\phi_q>_{sq})^{1/2} \tag{12}$$

The most common assumption used to justify Eq. (12) is that the subsystem "r" is large and unperturbed by the presence or absence of interactions with subsystem "s". Any differences among the various matrix elements, $<\psi_{q'}|V_{rsq}|\psi_q>_{rq}$, can be ignored. The external potentials emanating from the subsystem "r" are equal to each other up to a constant, for all q. The off-diagonal elements vanish. The weak-interaction-limit case permits the approximation,

$$V_{rsq}^{ext} \approx \text{constant} \approx <\psi_q|V_{rsq}|\psi_q>_{rq} . \tag{13}$$

The desired form for the ensemble average is therefore achieved. Eq. (13) suggests another approximation. The constant may depend on q. A second, less-restrictive approximation ignores the off-diagonal contributions in Eq. (12), which leads to the form

$$V_{rsq}^{ext} = <\psi_q|V_{rsq}|\psi_q>_{rq} . \tag{14}$$

A third approximation simply defines an interaction contribution to the external potential as the ratio

$$V_{rsq}^{ext} = <\Psi|V_{rsq}|\psi_q>_{rq}/<\Psi|\psi_q>_{rq} . \tag{15}$$

Whether or not either approach is applicable to any particular system must be determined from a separate investigation of $V_{rsq}$. If one cannot justify an assumption of this kind, then ensemble averaging and ensemble DFT will not be applicable. From here forward, we will assume that one of the Eqs. (12)-(15) is justifiable, so that we have the approximation

$$E \approx \left(\sum_q w_q (E_{rq} + <\phi_q|H_{sq} + V_{rsq}^{ext}|\phi_q>_{sq}/<\phi_q|\phi_q>_{sq})\right)/\left(\sum_q w_q\right) . \tag{16}$$

While we have allowed for approximations in reaching this form for an ensemble average, we have not restricted ourselves to the weak-interaction limit as is done in traditional statistical mechanics.

We take the approximate nature of Eq. (16) as a given, since we are dealing with situations where our knowledge of the total system is incomplete. The main difference between Eq. (16) and more traditional results is that the effective external potential is permitted to change with the partitioning of electrons between subsystems and the state "q" of the subsystem "r". This is necessary in order to make the states of "s" different from their isolated system counter parts represented by the hamiltonian $H_{sq}$. An interesting consequence to this fact is that the same partitioning of the electrons can have several external potentials associated with it, depending on the state q of the subsystem $\psi_q$ associated with the subsystem "r".

Rearranging terms we define the time-averaged subsystem energy $E_s$ by the familiar expression

$$E_s = \left(\sum_q w_q (E - E_{rq})\right)/\left(\sum_q w_q\right) = \left(\sum_q w_q E_{sq}\right)/\left(\sum_q w_q\right)$$

$$= \left(\sum_q w_q <\phi_q|H_{sq} + V_{rsq}^{ext}|\phi_q>_{sq}/<\phi_q|\phi_q>_{sq}\right)/\left(\sum_q w_q\right) . \qquad (17)$$

By analogy to our treatment of the energy, we define the time-averaged subsystem electron number $N_s$ by

$$N_s = \left(\sum_q w_q (N - N_{rq})\right)/\left(\sum_q w_q\right) = \left(\sum_q w_q N_{sq}\right)/\left(\sum_q w_q\right) . \qquad (18)$$

Note again that, by construction, each "rq" or "sq" subsystem state contains an integer number of electrons. This completes the basic derivation.

To this point no restrictions on which states of subsystem "r" have been included in the above derivation. One might think that, in order for DFT to be applicable to an ensemble state, only grounds states of $H_{rq}$ should be included. However the derivation allows any state where $V_{rsq}^{ext}$ is unique from the others, so that the associated subsystem wavefunction $\phi_q$ can be a ground state wave function. This argument applies to symmetry-defined uniqueness as well. Also note that we have not assumed that the $N_{rq}$ are unique. This allows for the possibility that the two different partitioning of the electrons might result in different effective external potentials representing the interaction between "r" and "s". Referring to our baseline example of a plasmon in a metal, consider the case when the electron of the electron-hole pair is in the reservoir. One would expect the effective external potential for the state where the electron is localized to be different from the effective external potential when that electron is delocalized. Whether or not this difference is appreciable is a separate issue to be decided from other information, and the availability of that information is yet another issue.

Next we examine the basic properties of open system density functional theory. In what follows, it will be assumed that all of the conditions necessary for the applicability of DFT have been met. Which states of subsystem "r" to include in Eq. (17) is not yet decided. One might think that, in order for DFT to be applicable to the ensemble state, only grounds states of $H_{rq}$ should be included. However, it will be shown that this is not necessarily the case. In addition, it will be shown that each integer-state contribution to the ensemble average is an integer-state density-functional in the sense of Levy[7] and Valone.[8]

## III. Open-system density functionals with variable external potentials

PPLB showed that energy density functional theory (DFT) can be generalized to accommodate the open system point of view.[3] By way of both review and motivation, we restate the general results of that and related work. The founding realization is that constrained search DFT[1,8] can be extended to a search over general, ensemble density matrices representing mixed states, in terms of particle numbers, of the system. This realization is extended further to the case where the external potentials for each integer-state in the ensemble can be different from each other. The central result of this Section is that the integer-state energies correspond to Levy[7] or Levy-Valone[8] functionals. This result will be proven through a pinching argument.

For the general case, suppose an open system with electron density, $n_s(x)$, is composed of integer-state electron densities, $n_{sq}(x)$,

$$n_{sq}(x) = N_{sq} \int dx_2 \ldots dx_{N_{sq}} \Gamma_{sq}(x, x_2, \ldots x_{N_{sq}}; x, x_2, \ldots x_{N_{sq}}) \tag{19}$$

and

$$n_s(x) = \sum_q W_q n_{sq}(x) , \tag{20}$$

with $N_{sq}$ integer numbers of electrons in the $q^{\underline{th}}$ distribution. Here it is simpler to use the probabilities defined in Eq. (4) and a normalized representation of $\Gamma_{sq}$. The $\Gamma_{sq}$ themselves may be either pure-states or an ensemble of states all having the same integer number of electrons for a given "sq". In terms of the previous Section, $\Gamma_{sq}$ is an average over subsystem wavefunctions, $\phi_q$, all of which have $N_{sq}$ electrons and the same interaction potential, $V_{rsq}{}^{ext}$.

The corresponding mixed-state density matrix $\Gamma_s$ is then given by

$$\Gamma_s = \Sigma_q \, W_q \, \Gamma_{sq} \, . \tag{21}$$

Following Perdew[10] we define an open-ensemble energy-density functional as

$$E_s[n_s] = \min \, \{<H_s, \Gamma_s>: \text{all } \Gamma_s \text{ reducing to } n_s\} \, , \tag{22}$$

for a trial time-averaged electron density $n_s$ of the subsystem. The expectation value means precisely the form of the ensemble average in Eq. (15):

$$<H_s, \Gamma_s> = \Sigma_q \, W_q \, <H_{sq} + V_{rsq}{}^{ext}, \Gamma_{sq}> \, . \tag{23}$$

(Note that the subscript "sq" denoting the integration space has been dropped. It is no longer needed for clarity.) Because of the definitions chosen to this point,

$$<\Gamma_s> = \Sigma_q \, W_q \, <\Gamma_{sq}> = 1 \, , \tag{24}$$

as required. The interpretation of the quantities in Eqs. (22) and (23) is crucial. A choice for $n_s(x)$ means that the integer-states contributing to the ensemble average have been chosen, an electron density, $n_{sq}$, for each integer-state has been chosen, and an occupation number, $W_q$, for each integer-state has been chosen. Also, expressed in this way, there is no need for lagrange multipliers to enforce constraints in performing the minimization in Eq. (22). Also, a choice for $n_s(x)$ also determines $N_s$. The main result can now be proven, namely, that the integer-state term in the ensemble average must be an integer-state energy-density functional. Because the $W_q$'s are fixed by ones choice of $n_s(x)$ and are positive, the minimum of Eq. (22) must be bounded below by the minimum value of each term in Eq. (23):

$$\min \{<H_s, \Gamma_s>: \text{ all } \Gamma_s \text{ reducing to } n_s\} \geq$$

$$\Sigma_q W_q \min \{<H_{sq} + V_{rsq}^{ext}, \Gamma_{sq}>: \text{ all } \Gamma_{sq} \text{ reducing to } n_{sq}\} . \qquad (25)$$

The minimum of each individual term is just a Levy-Valone density energy functional,[7,8]

$$\min \{<H_{sq} + V_{rsq}^{ext}, \Gamma_{sq}>: \text{ all } \Gamma_{sq} \text{ reducing to } n_{sq}\} = \mathbf{E}[n_{sq}] ;$$

$$\mathbf{E}[n_{sq}] \equiv <H_{sq} + V_{rsq}^{ext}, \Gamma_{sq}^{min}> , \qquad (26)$$

for $n_{sq}(x)$ for the system with the external potential modified by $V_{rsq}^{ext}$, which comes from averaging over the coordinates of subsystem "rq". The integer-state density matrix that achieve the minimum in Eq. (26) are designated as $\Gamma_{sq}^{min}$. These integer-states constitute the components of a perfectly acceptable trial $\Gamma_s$ in Eqs. (22) and (23). That is, suppose that

$$\Gamma_{sq}^{trial} = \Sigma_q W_q^{trial} \Gamma_{sq}^{min} . \qquad (27)$$

This means that the $\Gamma_{sq}^{min}$ must also form an upper bound to Eq. (22), for the chosen $W_q^{trial}$. These two bounding arguments force one to conclude that

$$\Sigma_q\, W_q^{trial}\, \mathbf{E}[n_{sq}] \geq \min\, \{<H_s, \Gamma_s>: \text{all } \Gamma_s \text{ reducing to } n_s\}$$

$$\geq \Sigma_q\, W_q^{trial}\, \mathbf{E}[n_{sq}]\, . \tag{28}$$

Consequently, in this construction, an open-system energy-density functional is precisely an average over integer-state density functionals. No assumptions have been made about the relationship between the occupation numbers, $W_q$, and the average particle-number. The only properties that are require to prove Eq. (28) are that the $W_q$ are positive and fixed for a given trial ensemble density matrix.

In the next Section, we shall see that the integer-state functionals inherit two important properties through this relationship. The first is that the integer-state functionals inherit chemical potential equalization whenever it applies among the subsystems. The second is that the integer-state functionals inherit the time-superposition average of chemical potentials when there are multiple, competing processes between subsystems. That is, if it is possible for the subsystem "s" to both gain and loose electrons to "r", then the chemical potential governing the behavior of the integer-state functionals is the time-superposition of the individual processes.

## IV. Properties of Open-System Density Functionals

In this section, a slightly different approach is followed in making the association between changes in the subsystem energy, $E_s$, with changes in the average number of particles in the system, $N_s$. This approach is followed because it is straightforward, it obviates the connection between finite-difference relationships for the chemical potential and functional-derivative relationships, and it leads to a Gibbs-Duhem-like equation for the integer-state functionals. From Eqs. (1) - (3), one has the option of solving for any one of the occupation numbers, say $W_0$, and rewriting Eq. (2) as

$$N_s = N_{s0} + \sum_{q \neq 0} W_q (N_{sq} - N_{s0}) . \tag{29}$$

Similarly, for the ensemble energy,

$$E_s = E_{s0} + \sum_{q \neq 0} W_q (E_{sq} - E_{s0}) . \tag{30}$$

To further the analogy between DFT and classical thermodynamics, the occupation numbers are rescaled to charge-transfer coefficients, $C_q$. Specifically, these may be defined by the relationship

$$C_q = W_q (N_{sq} - N_{s0}) \equiv W_q \Delta N_q , \tag{31}$$

whereby the electron-number constraint becomes

$$N_s = N_{s0} + \sum_{q \neq 0} C_q, \tag{32}$$

and the energy as a function of the $C_q$ takes the form

$$E_s = E_{s0} + \sum_{q \neq 0} C_q \Delta E_q / \Delta N_q , \tag{33}$$

where $\Delta E_q \equiv E_{sq} - E_{s0}$. The ratios $\Delta E_q/\Delta N_q$ conform identically to the operational definition of specific free energies of classical thermodynamics.[11] Therefore we are perfectly well justified in identifying them as chemical potentials, $\mu_q$, and expressing the total energy as

$$E_s = E_{s0} + \Sigma_{q \neq 0} \, C_q \, \mu_q \, , \qquad (34)$$

It is plain to see now that an analog of the Gibbs free-energy does exist[11] for open-system DFT. The next topic uses this expression to make the identification between the variation of the total energy with respect to the average number of electrons and its variations with respect to integer-state electron densities.

Now we want to minimize the energy subject to the constraint that the $N_s$ has a prescribed value. Using the lagrange multiplier method, we may construct the auxiliary functional $\Omega$ such that

$$\Omega = E_s - \mu \, (N_{s0} + \Sigma_{q \neq 0} \, C_q) \, . \qquad (35)$$

Calculating derivatives with respect to any of the $C_q$ for fixed densities, $n_{sq}$, yields

$$\partial \Omega / \partial C_q = \Delta E_q / \Delta N_q - \mu \, . \qquad (36)$$

Setting each of these partial derivatives to zero yields the relationships

$$\mu = \Delta E_1/\Delta N_1 = \ldots = \Delta E_S/\Delta N_S \, , \qquad (37)$$

for all states, 1 through S, contributing to the mixed state. That is, the chemical potential of the system is the chemical potential on a state-by-state basis. This conclusion has been reached independent of whether the chemical potential is that appropriate to the subsystem "s" or is the global chemical potential of a large, immutable reservoir. See Reif for a discussion of this distinction.[5] Finally, because independent variations may be combined arbitrarily, it is permissible to take a variation of $\Omega$ in the collective variable $N_{s0} + \Sigma_{q \neq 0} C_q = N_s$ to conclude that

$$\partial\Omega/\partial N_s = 0 = \partial E_s/\partial N_s - \mu . \tag{38}$$

Thus, we find, in agreement with others,[10-12] identification the derivative of the energy with respect to the time-average number of electrons, which is a continuous variable, with the chemical potential. The same conclusions is reached if the one starts the weights as in Eqs. (17) and (18), instead of the occupation numbers.

At the same time, for Eq. (37) to be true, integer-state electron densities, $n_{sq}$, must have chosen to make the energy stationary under the electron-number constraint. In a trivial manipulation, for each and every q,

$$\Delta E_q = \mathbf{E}[n_{sq}] - \mathbf{E}[n_{s0}] = \mu \, \Delta N_{sq} = \mu \, (\mathbf{N}[n_{sq}] - \mathbf{N}[n_{s0}]) , \tag{39}$$

$\mathbf{E}[n_{sq}]$ being the integer-state energy-density functional, Eq. (26), as defined by constrained-search theory[7,8] and $\mathbf{N}[n_{sq}]$ being the norm operator, but the nontrivial outcome is that the functional derivative with respect to any one $n_{sq}$ or $n_{s0}$ must satisfy the relationship

$$\delta\mathbf{E}[n_{sq}]/\delta n_{sq} = \mu . \tag{40}$$

This establishes the fundamental connection between the finite-difference forms for the chemical potential and the functional-derivative forms. In addition, other known relationships follow from the above. The relationship between the functional derivative of $E_s$ with respect to an integer-state density and the integer-state functional derivative itself is

$$\delta E_s[n_s]/\delta n_{sq} = W_q\, \delta \mathbf{E}[n_{sq}]/\delta n_{sq} = W_q\, \mu\,, \qquad (41)$$

for any q. The result is true for q = 0 as well, but some rearrangement is needed to get that term in the same form as the others. Through this result we can see that there is some need to distinguish between functional derivatives of the time-averaged and the integer-state energies. As in Parr et al.,[12] we find that

$$\sum_q \langle n_{sq}\, \delta E_s[n_s]/\delta n_{sq}\rangle = \sum_q W_q \langle n_{sq}\, \delta \mathbf{E}[n_{sq}]/\delta n_{sq}\rangle = \mu\, N_s\,. \qquad (42)$$

These results comprise the essential integer-state functional derivative properties.

Next consider the much more highly constrained case where the occupation numbers are determined thermodynamically. Then the $w_q$ are functionals of the $\mathbf{N}_{sq}$ and the $\mathbf{E}_{sq}$ as in Eq. (5). First we show that the $\mathbf{E}_{sq}$ are still Levy-Valone functionals. The same conclusion regarding state-by-state chemical potential equalization in the non-thermodynamic case follow for the thermodynamic case as well. Then we will show that the connection between finite-difference forms and functional derivatives also remains valid, up to a term equal to $k_B T$.

The arguments follow the same pinching strategy as developed in Eqs. (25) - (28). The upper-bound argument of Eqs. (27) and (28) is still valid. The lower bound argument requires a slight modification. When the $\Gamma_{sq}$ are varied in the constrained search procedure, the $w_q = \exp((\mu N_{sq} - E_{sq})/k_B T)$ vary as well. Under the assumption that the integer-states are bound states, i. e. the $E_{sq} < 0$, and much weaker assumption that the $w_q \geq 0$, the minimum value of each term in Eq. (23) is bounded below by

$$\min \{<H_s, \Gamma_s>: \text{ all } \Gamma_s \text{ reducing to } n_s\} \geq$$

$$\Sigma_q \max \{W_q : \text{ all } \Gamma_{sq} \text{ reducing to } n_{sq}\}$$

$$\bullet \min \{<H_{sq} + V_{rsq}^{ext}, \Gamma_{sq}>: \text{ all } \Gamma_{sq} \text{ reducing to } n_{sq}\} . \tag{43}$$

The maximum values of the $W_q$ are be bound below by the $E_{sq}$ and $N_{sq}$ derived from any trial $\Gamma_s$:

$$\max \{W_q : \text{ all } \Gamma_{sq} \text{ reducing to } n_{sq}\} \geq W_q(\{E_{sq}[\Gamma_{sq}^{trial}], N_{sq}[\Gamma_{sq}^{trial}]\}_{\text{all } q}) . \tag{44}$$

Again, a valid choice for $\Gamma_s$ is the one yielding the Levy-Valone energy. Set the $E_{sq}$ to those energies,

$$E_{sq}[\Gamma_{sq}^{trial}] = E_{sq}[\Gamma_{sq}^{min}] = \mathbf{E}_{sq}[n_{sq}] . \tag{45}$$

The pinching argument provides the bounds,

$$\Sigma_q W_q(\{\mathbf{E}_{sq}[n_{sq}], \mathbf{N}_{sq}[n_{sq}]\}_{\text{all } q}) \mathbf{E}[n_{sq}] \geq \min \{<H_s, \Gamma_s>: \text{ all } \Gamma_s \text{ reducing to } n_s\}$$

$$\geq \Sigma_q W_q(\{\mathbf{E}_{sq}[n_{sq}], \mathbf{N}_{sq}[n_{sq}]\}_{\text{all } q}) . \tag{46}$$

Thus, the integer-state energy-density functionals are again shown to be the Levy-Valone functionals.[7,8] The individual weights can be though of as density functionals as well:

$$w_q[n_{sq}] = \exp((\mu \mathbf{N}_{sq}[n_{sq}] - \mathbf{E}_{sq}[n_{sq}])/k_B T) , \tag{47}$$

Viewing the thermodynamic case as a subset of all the possible ensemble averages defined above, these results are sensible ones. Note also that the actual form of the $w_q$ is not required for Eq. (46) to be true.

Taking functional derivatives of the auxillary functional,

$$\Omega = \left(\sum_q w_q[n_{sq}] \left(\mathbf{E}_{sq}[n_{sq}] - \mu \mathbf{N}_{sq}[n_{sq}]\right)\right) / \left(\sum_q w_q[n_{sq}]\right) , \tag{48}$$

requires taking functional derivatives of the weights as well. For any particular weight,

$$\delta w_q[n_{sq}]/\delta n_{sq} = \beta \left(- \mu\, \delta \mathbf{N}_{sq}[n_{sq}]/\delta n_{sq} - \delta \mathbf{E}_{sq}[n_{sq}]/\delta n_{sq}\right) w_q . \tag{49}$$

where $\beta = 1/k_B T$. For $\delta\Omega/\delta n_{sq} = 0$, one obtains

$$0 = w_q \left(\delta \mathbf{E}_{sq}[n_{sq}]/\delta n_{sq} - \mu\, \delta \mathbf{N}_{sq}[n_{sq}]/\delta n_{sq}\right)$$

$$\left(1 - \beta \left(\mathbf{E}_{sq}[n_{sq}] - E - \mu \left(\mathbf{N}[n_{sq}] - N\right)\right)\right) . \tag{50}$$

Any of the three terms can be zero. If $w_q = 0$, the state q is unoccupied and can be excluded from the ensemble. The second term gives a slightly different result,

$$\mathbf{E}[n_{sq}] - E = k_B T + \mu \left(\mathbf{N}[n_{sq}] - N\right) , \tag{51}$$

The finite-difference form in the thermodynamic case differs from non-thermodynamic form, Eq. (39), by $k_B T$. The third term gives the customary result of Eq. (40). Taking the functional derivative of Eq. (51) with respect to $n_{sq}$, one again finds consistency between the finite-difference forms and the functional derivative forms.

The last key result is related to PPLB and other work concerning the continuity of the functional with respect to functional derivatives that cross through an integer state. In that spirit consider the three systems discussed after Eq. (12) of PPLB, the slightly positive, slightly negative and the exactly neutral cases. Consider also that the system of interest is an atom that is only very weakly interacting with a reservoir and is very near 0 K. Then PPLB argue on physical grounds that the chemical potential of the slightly positive atom must be $\mu_+ = -I$, the negative of the ionization potential, the slightly negative atom, $\mu_- = -A$, the negative of the electron affinity and the neutral, $\mu_0 = -1/2 \, (I+A)$, the negative of the electronegativity. The chemical potential may exhibit discontinuous behavior as a function of the charge on the system, in the cases where the three chemical potentials are not all equal. This jump discontinuity appears as discontinuity in the functional derivative of the time-average or ensemble-average energy-density functional.

We want to know how this derivative discontinuity manifests itself in the functional derivatives of the integer states as a function of charge on the system. Consider the discussion from PPLB where a system consisting of a single atom is placed in contact with a reservoir at low temperature. The reservoir is allowed to change in such a way that the charge on the atom varies from positive to neutral to negative. The corresponding ensemble averages for the positive and negative states are:

$$N_+ = \mathbf{N}_0 + W_+ \, (\mathbf{N}_+ - \mathbf{N}_0) \; ; \; N_- = \mathbf{N}_0 + W_- \, (\mathbf{N}_- - \mathbf{N}_0) \, . \tag{52}$$

and

$$E_+ = \mathbf{E}_0 + W_+ \, (\mathbf{E}_+ - \mathbf{E}_0) \; ; \; E_- = \mathbf{E}_0 + W_- \, (\mathbf{E}_- - \mathbf{E}_0) \, . \tag{53}$$

Of course, $\mathbf{N_+} - \mathbf{N_0} = -1$ and $\mathbf{N_-} - \mathbf{N_0} = +1$, but the development above requires that the functional notion be kept for clarity. The occupation numbers have the thermodynamic form of Eq. (47). Because the average particle-number is specified,[4,5] the chemical potentials are determined as a function of the $N_s$. From Eq. (40), in the slightly positive case,

$$\delta \mathbf{E_+}[n]/\delta n \big|_{n=n_0} = \mu_+ , \qquad (54)$$

and in the slightly negative case,

$$\delta \mathbf{E_-}[n]/\delta n \big|_{n=n_0} = \mu_- . \qquad (55)$$

Now note that the time-averaged energy of the neutral state is the superposition of the time-averaged positive and negative state energies:

$$E_0 = 1/2\, (E_+ + E_-) , \qquad (56)$$

and

$$\delta \mathbf{E}[n]/\delta n \big|_{n=n_0} = \mu_0 . \qquad (57)$$

Physical consistency suggests that functional derivatives of these processes obey time-superposition as well. This consistency requires that

$$\mu_0 = 1/2\, (\mu_+ + \mu_-) . \qquad (58)$$

In this derivation, the factor of one half originates in the same factor in Eq. (56), which in turn is required by neutrality, $N_0 = 1/2\, (N_+ + N_-)$ and $W_+ = W_-$. It is not arbitrary in this analysis.

Moreover, if these three chemical potentials are not all equal, the integer-state functionals inherit the derivative discontinuities of the ensemble energy. For the three cases considered in PPLB, the functional derivative with respect to the neutral integer-state density can have any of the chemical potentials in Eqs. (54), (55), and (57). However, these are Levy or Levy-Valone functionals cannot have a derivative discontinuity in the exchange-correlation functional. (Otherwise, universality would be violated.) In each case, the neutral state density distorts discontinuously so as to satisfy the condition imposed by the specified charge for that case. The chemical potential acts as a kind of boundary condition on the density which is most easily understood as governing its long-range behavior.[2,3,13,14,15] Notice though that the issue of whether or not there is a solution exists to the variational equation for a particular value of $\mu$ and **N** is separate issue from this investigation.

The boundary condition on the long-range behavior is intimately connected to the long-range behavior of the external potential of the system. In the ensemble-average construction presented here, the external potentials can differ in their long-range behavior, as they need not tend to the same zero of energy.[16] The external potentials of the ensemble averages differ from their purely isolated-subsystem form by the interactions between the subsystem and other subsystems or a reservoir. The nature of the subsystem-subsystem(reservoir) interaction includes exchange-correlation contributions. In the parlance established here, the exchange-correlation contributions are embedded in the effective external potentials.

A more general, three-system case can also be described. Consider a generalized form of Eq. (56), $E = \alpha E_+ + (1-\alpha) E_-$, where $\alpha \geq 0$. How does the chemical potential vary with $N$ when all three states, +, 0, and –, are allowed to participate, even if $N - N_0$ is positive or negative. First, redefine the integer-state occupation numbers for the three-sate system to read

$$E = \mathbf{E}_0 + \alpha W (\mathbf{E}_+ - \mathbf{E}_0) + (1-\alpha) W (\mathbf{E}_- - \mathbf{E}_0) . \tag{59}$$

Then it must be that

$$W = W_+/\alpha = W_-/(1-\alpha) . \tag{60}$$

Solving for $W_- = N - N_0 + W_+$, one finds for $\alpha$,

$$\alpha = W_+/(N - N_0 + 2 W_+) . \tag{61}$$

It is convenient to define the variable $a = (N - N_0)/W_+$. The generalization of Eq. (58) is that

$$\mu(a) = (1/(a+2)) \mu_+ + ((a+1)/(a+2)) \mu_- . \tag{62}$$

Now consider the following three cases in which all three states are assumed to participate. Because all three states are assumed to be occupied, $-1/2 \leq N - N_0 \leq 1/2$ and $0 \leq W_+ \leq 1/2$. First, suppose that $N - N_0 \rightarrow -1/2$, the cationic case. Constraints on the occupation numbers require that $W_+ \rightarrow +1/2$, so that $a \rightarrow -1$. Then, $\mu \rightarrow \mu_+$ and the anionic state becomes unoccupied, as is necessary physically. Second, suppose that the subsystem interactions are allowed to become such that $N - N_0 \rightarrow 0$, the neutral case. Then, $\mu \rightarrow \mu_0$, in agreement with Eq. (58). First, suppose that $N - N_0 \rightarrow +1/2$, the cationic case. Constraints on the occupation numbers require that $W_- \to +1/2$, so that $a \to +\infty$. Then, $\mu \to \mu_-$ and the cationic state becomes unoccupied, as is necessary physically. In this scenario, the presence of the all three states in the ensemble average allows a continuous transition among the chemical potentials of the underlying time-averaged processes. Certainly this scenario is not an equilibrium thermodynamic one, but physically plausible in a general sense.

Finally, consider the definition of the interaction potential defined in Eq. (12). The partitioning of the bare interactions, $V_{rsq}$, between subsystems is somewhat arbitrary. As originally defined, all of the interaction is applied to "s" and none to the rest of the system. $V_{rsq}$ could be partitioned between "r" and "s" in a way which makes the ensemble approximation as close as possible E. Then, one can consider an self-consistency procedure in the spirit of Rychlewski and Parr. A partition of $V_{rsq}$ is chosen. Approximate wavefunctions for "r" are chosen or calculated. The $n_{sq}$ (or wavefunctions) for "s" are solved for as described above. In turn, these can be used calculate a new effective interaction for "r" and those densities or wavefunctions refined. The process would be continued until the desired level of self-consistency is achieved.

## V. Summary

In summary, the energy functionals of the integer-states composing an ensemble energy-density functional are shown to be Levy-Valone functionals.[7,8] The principle of chemical potential equalization has been shown to apply to the individual integer-states in an ensemble average which is at its ground state.  A clear routine to equating finite-difference forms of the chemical potential to functional derivative forms has been identified.  This route also yields a form for the ensemble energy which resembles a Gibbs-Duhem expression.  The results for the case where the ensemble is a thermodynamic one are consistent with the general case.  The results also apply when the effective external potentials of the integer-states of the ensemble, as determined by interactions between subsystems or between a subsystem and a reservoir, differ by nonconstant values.  Finally, the derivative discontinuity of the time-averaged or ensemble-averaged energy density may operate more like a discontinuity in the boundary conditions of the integer-state functionals than a discontinuity in the exchange-correlation functional itself.  The boundary conditions are related to the long-range behavior of the effective external potentials which need not tend to the same zero of energy.

The implications for these results have a bearing on how energy-density functional calculations are performed, the construction of empirical potential energy surfaces, and the construction of new energy-density functionals.

## Acknowledgment

I wish to thank Professor Susan Atlas, Department of Physics and Astronomy, University of New Mexico for numerous invaluable discussions and suggestions.  This work was performed at Los Alamos National Laboratory under the auspices of the U. S.

Department of Energy, under contract No. W-7405-ENG-36, and funded in part through its Center for Semiconductor Modeling and Simulation, a CRADA program performed jointly with the Semiconductor Research Corporation.